\documentclass[twocolumn]{revtex4}

\usepackage{amsmath,amssymb}
\usepackage{graphicx,psfrag,float,epsfig,color}
\usepackage{mathrsfs,wasysym}
\usepackage{epstopdf}
\usepackage{enumitem}

\DeclareSymbolFontAlphabet{\mathrsfs}{rsfs}
\DeclareMathAlphabet{\mathcal}{OMS}{cmsy}{m}{n}

\newcommand{\be}{\begin{equation}}
\newcommand{\ee}{\end{equation}}

\begin{document}


\title{Null infinity waveforms from extreme-mass-ratio inspirals in Kerr spacetime}

\author{An{\i}l Zengino\u{g}lu$^1$} 
\author{Gaurav Khanna$^2$}
 
\affiliation{$^1$Theoretical Astrophysics, California Institute of Technology, Pasadena, California USA} 

\affiliation{$^2$Department of Physics, University of Massachusetts, 
Dartmouth, Massachusetts USA}

\begin{abstract}
We describe the hyperboloidal compactification for Teukolsky equations in Kerr spacetime. We include null infinity on the numerical grid by attaching a hyperboloidal layer to a compact domain surrounding the rotating black hole and the orbit of an inspiralling point particle. This technique allows us to study, for the first time, gravitational waveforms from large- and extreme-mass-ratio inspirals in Kerr spacetime extracted at null infinity. Tests and comparisons of our results with previous calculations establish the accuracy and efficiency of the hyperboloidal layer method.
\end{abstract}


\maketitle

\section{Introduction}

Black hole perturbation theory plays a prominent role in obtaining physical insight into the quantitative behavior of astrophysical systems \cite{Regge:1957td, Zerilli:1970se, Bardeen:1973xb, Teukolsky:1972my, Teukolsky:1973ha, Chandrasekhar:1992bo, Kokkotas:1999bd, Nagar:2005ea}. In recent years there has been considerable interest in computing the gravitational waveforms emitted by the radiation-reaction driven inspiral of a small, compact object into a large supermassive black hole. The waveforms from such extreme-mass-ratio inspirals (EMRIs) are expected to be observed by future, space-based, gravitational wave detectors such as the Laser Interferometric Space Antenna. 

EMRIs provide a rich phenomenology relevant for astrophysical and theoretical questions. Observations of EMRIs would deliver, among others, an accurate mapping of black hole spacetimes \cite{Hughes:2000ssa} and tests of the black hole uniqueness theorem \cite{Ryan:1997hg}.
On the theoretical side, modeling of EMRIs poses challenging problems. Because the numerical solution of full Einstein equations for extreme mass ratios of the order of $10^{-5}$ are computationally prohibitive, EMRIs are calculated via approximation techniques. Currently, the most common approximative methods to study EMRIs are the kludge waveform approach \cite{Babak:2006uv, Gair:2005ih, Sundararajan:2007jg, Sundararajan:2008zm, Sundararajan:2010sr}, the effective-one-body (EOB) approach \cite{Buonanno:1998gg, Buonanno:2000ef, Damour:2000we, Nagar:2006xv,Damour:2007xr, Yunes:2009ef, Yunes:2010zj, Bernuzzi:2010ty}, and the self-force approach \cite{Quinn:1996am, Mino:1996nk, Detweiler:2002mi, Barack:2007tm, Canizares:2010yx, Barausse:2010ka, Dolan:2011dx}.

In these methods, the waveforms can be calculated by numerically solving the equations describing black hole perturbations. When the central, supermassive black hole is assumed to be rotating, the most common approach is to solve the Teukolsky equations that describe curvature perturbations of a Kerr spacetime \cite{Teukolsky:1972my,Teukolsky:1973ha}.

A difficulty in solving the Teukolsky equations numerically---especially in the time domain---is the construction of suitable outer boundary conditions. Typically, the computational domain is artificially truncated to calculate the solution in a finite domain. One needs to provide boundary data along the timelike boundary of this domain such that the artificial boundary is transparent to outgoing waves from the interior. The outer boundary problem is already nontrivial for the evolution of scalar waves in Minkowski spacetime \cite{GroteKeller, Alpert00, HagstromLau}. There has been substantial progress in constructing good outer boundary conditions for wave equations in Schwarzschild spacetime \cite{Lau:2004as, Lau:2004jn, Buchman:2006xf, Buchman:2007pj}, which have been recently generalized to Kerr spacetime \cite{Deadman:2011zz}. Such sophisticated boundary conditions, however, can be difficult to implement. To our knowledge, there is no numerical implementation of the improved boundary conditions in Kerr spacetime.

Another difficulty in the numerical solution of the Teukolsky equation is the extraction of gravitational waves as measured by idealized observers at null infinity. Astrophysical sources of gravitational radiation are typically thousands of light years away, whereas the computational truncation is performed at a few thousand Schwarzschild radii. One argues that the dynamics in the asymptotic domain is negligible. However, there are certain effects, such as the backscattering off curvature \cite{Price:1971fb,Gundlach:1993tp}, where asymptotic properties of the solution are essentially different from the corresponding properties at finite distances. Such differences can be relevant to gravitational wave astronomy. For example, it has been found that the asymptotic formula relating the curvature perturbation $\psi_4$ to the gravitational wave strain (Eq.~\eqref{eq:psi4hphm}) is invalid for polynomially decaying solutions even at large distances used in numerical wave extraction \cite{Zenginoglu:2009ey}.  Further, we know that the phase of the gravitational wave signal measured at null infinity and at finite distances may differ substantially in EMRIs \cite{Bernuzzi:2010xj, Bernuzzi:2011aj}.

A clean solution to both the outer boundary and the radiation extraction problems is to include null infinity in the computational domain. No artificial boundary conditions are needed when the solution is calculated globally in space, and the idealized gravitational wave signal can be read off directly at null infinity.

There are two methods to include null infinity in the computational domain: the characteristic method and the hyperboloidal method. These methods correspond to approaching null infinity along null and spacelike directions respectively. 

The characteristic method is very well developed \cite{lrr-2009-3}. It has been applied recently to the unambiguous computation of gravitational waves from binary black hole mergers simulated via Einstein equations \cite{Reisswig:2009us, Reisswig:2009vc, Babiuc:2011qi}. Concerning the Teukolsky equation, it has been used in Schwarzschild spacetime for black hole collisions in the close limit \cite{Campanelli:2000in,Husa:2001pk}. The construction of a null foliation in Kerr spacetime, however, is rather complicated \cite{Pretorius:1998sf, FletcherLun, Venter:2005cs, Bai:2007rs}. As a result, there is no numerical computation of gravitational perturbations of Kerr spacetime at null infinity. 

The second approach to including null infinity in the computational domain is the hyperboloidal approach \cite{Friedrich83}. The construction of suitable hyperboloidal foliations is simpler than characteristic ones because spacelike surfaces are more flexible than null surfaces \cite{Zenginoglu:2007jw}. Various studies have demonstrated the accuracy and efficiency of the hyperboloidal method in dealing with perturbations of black hole spacetimes, but most of them are restricted to spherical symmetry. For example, numerical studies have been performed on Yang-Mills equations in a Schwarzschild spacetime  \cite{Zenginoglu:2008wc, Bizon:2010mp}, spherically symmetric wormholes \cite{Gonzalez:2009hn} and boson stars \cite{LoraClavijo:2010xc}, self-force computations \cite{Vega:2011wf} and gravitational perturbations \cite{Zenginoglu:2008uc,Zenginoglu:2009ey} including extreme-mass-ratio inspirals \cite{Bernuzzi:2010xj, Bernuzzi:2011aj}. 

The only numerical computations of black hole perturbations using the hyperboloidal method without spherical symmetry deal with scalar perturbations \cite{Zenginoglu:2009hd, Zenginoglu:2010zm, Racz:2011qu, Jasiulek:2011ce}. One goal of this paper is to present the application of the hyperboloidal method to Teukolsky equations in Kerr spacetime for the calculation of gravitational waveforms at null infinity. The test problem we study to demonstrate the method is of astrophysical interest. The gravitational waveforms we compute are emitted by the perturbations due to a point particle inspiralling into the central Kerr black hole. 

Another goal of this paper is to solve a difficulty of applying the hyperboloidal method in Kerr spacetime. The hyperboloidal foliation of Kerr spacetime constructed in \cite{Zenginoglu:2007jw} uses a transition zone in which a truncated Cauchy surface is matched to a hyperboloidal surface. This technique requires tuning a large number of foliation parameters and results in a sudden drop of characteristic speeds in the transition zone decreasing computational efficiency \cite{Zenginoglu:2009ey, Zenginoglu:2009hd, Vega:2011wf}. There are two solutions to this problem. One is to use a single smooth surface avoiding the transition zone of \cite{Zenginoglu:2007jw}. This technique has recently been applied by R\'acz and T\'oth in a detailed study of polynomial decay rates of a scalar field in Kerr spacetime \cite{Racz:2011qu}. They construct the first smooth, horizon-penetrating, hyperboloidal foliation of Kerr spacetime for their study. Modifying the coordinates everywhere smoothly leads to very efficient numerical computations. However, it may be favorable due to practical reasons to employ standard coordinates in the strong field region near the rotating black hole, especially when particles are present in the computational domain, so that the description of the particles' motion does not need to be changed.

In this paper we use the recently proposed hyperboloidal layer (\cite{Zenginoglu:2010cq}) instead of a transition zone. The new layer technique has been applied in studies of EMRIs using the Regge--Wheeler--Zerilli formalism completed by EOB-resummed analytical radiation reaction (at linear order in the mass ratio) \cite{Bernuzzi:2010xj, Bernuzzi:2011aj}. In this paper, we demonstrate the construction and application of a hyperboloidal layer for the Teukolsky formalism and the kludge approach in Kerr spacetime. 

The paper is organized as follows. In Section II we present the standard Teukolsky formalism with the transformations typically used in numerical calculations. In Section III we present how the hyperboloidal layer technique is applied to Kerr spacetime and to the Teukolsky equations as two simple coordinate transformations. Numerical results obtained with this formalism are presented in Section IV. After describing briefly the numerical method, we present a comparison of waveforms, an improved agreement of the energy fluxes with frequency-domain computations for several circular-equatorial orbits, a discussion of recoil velocities, and an example simulation that demonstrates remarkable gains in efficiency with the new method. A discussion of our results and an outlook can be found in Section V.

\section{The Teukolsky formalism}

Curvature perturbations of Kerr spacetime are governed by a separable equation in the Newman--Penrose formalism \cite{Teukolsky:1972my,Teukolsky:1973ha}. The original calculations involve the Boyer--Lindquist representation of the Kerr metric. It is convenient for numerical applications to introduce certain transformations of the Teukolsky equation. We review the common transformations to set up the equations to which the hyperboloidal method will be applied. 

\subsection{Teukolsky equation in Boyer--Lindquist coordinates}

The Kerr metric in Boyer--Lindquist coordinates $(t,r,\theta,\varphi)$ reads 
\begin{eqnarray} \label{bl_metric} 
&g_{\textrm{BL}}&= -\left(1-\frac{2Mr}{\Sigma}\right) dt^2 - \frac{4 a M r}{\Sigma}\sin^2\theta \,dt\,d\varphi + \frac{\Sigma}{\triangle}\, dr^2 \nonumber \\
&& \!\!\! +\Sigma \,d\theta^2+\left(r^2+a^2+\frac{2Ma^2r\sin^2\theta}{\Sigma}\right)\,\sin^2\theta\,d\varphi^2,\end{eqnarray}
where $\Sigma := r^2 + a^2 \cos\theta^2$, and $\triangle :=r^2 + a^2 - 2Mr$. We denote the mass of the Kerr spacetime by $M$, its angular momentum by $a M$. 

The Teukolsky equation describes curvature perturbations with spin weight $s$ in an adapted Newman-Penrose orthonormal frame. The homogeneous Teukolsky equation in Boyer--Lindquist coordinates reads
\begin{eqnarray*}
D^2 \partial_t^2 \Psi &=& - \frac{4aMr}{\triangle}\,\partial_\varphi\partial_t \Psi 
+ \triangle^{-s}\,\partial_r (\triangle^{s+1}\partial_r\Psi) \\
&+& \frac{1}{\sin\theta}\,\partial_\theta(\sin\theta\,\partial_\theta \Psi)
+ \left( \frac{1}{\sin^2\theta} - \frac{a^2}{\triangle}\right) \partial_\varphi^2\Psi \\
&-& 2s\left(\frac{M(a^2-r^2)}{\triangle}+(r+i a \cos\theta) \right) \partial_t\Psi \\
&+& 2s\left(\frac{a(r-M)}{\triangle}+\frac{i\cos\theta}{\sin^2\theta}\right)\partial_\varphi \Psi \\
&-& (s^2\cot^2\theta - s) \Psi,
\end{eqnarray*}
where $D^2 = (r^2+a^2)^2/\Delta - a^2\sin^2\theta$. In the following, we review the transformations that are typically applied to the Teukolsky equation for numerical computations \cite{Krivan:1997hc, PazosAvalos:2004rp, Sundararajan:2007jg}.

\subsection{Transformations of the Teukolsky equation}\label{trafos}

Boyer--Lindquist coordinates have undesirable features for numerical computations. Their time slices intersect at the bifurcation sphere leading to a singular radial coordinate at the horizon, in a similar way as Schwarzschild time slices do. There are two ways to deal with the coordinate singularity at the horizon: transform to a horizon-penetrating time slicing, or push the horizon to infinite coordinate distance with the tortoise coordinate. 

Horizon-penetrating coordinates are regular at the event horizon. They are typically used in combination with excision. The numerical treatment of the inner boundary is clean with this method because there are no characteristics coming out of the horizon. In addition, the radiation absorbed by the black hole can be calculated accurately. 

The most common approach in perturbation theory, however, is to use the tortoise coordinate. The tortoise coordinate pushes the horizon to infinite coordinate distance. The infinite domain is then truncated at a finite distance, where artificial ingoing boundary conditions are imposed. These lead to a contamination of the physical solution due to numerical reflections from the inner boundary.  But these reflections are small because the potential terms in the Teukolsky equation fall off exponentially fast in the tortoise coordinate, so the ingoing boundary conditions are quite accurate. In our study, we focus on the outgoing radiation and accept small reflections from the inner boundary. The truncation of the infinite domain at a finite coordinate implies that the efficiency of the code is not optimal because a large region in negative tortoise coordinate needs to be included in the computational domain. 

The relation of the tortoise coordinate to horizon-penetrating coordinates near the event horizon is similar to the relation of standard Cauchy coordinates to hyperboloidal coordinates near null infinity \cite{Zenginoglu:2009ey,Zenginoglu:2011jz}. A fundamental difference, however, is that the structure of the solution in the strong field region can be interesting, whereas the asymptotic solution is essentially simple in asymptotically flat spacetimes. The additional resolution that the tortoise coordinate provides near the black hole can be relevant in certain studies. This question should be analyzed in more detail. We choose to employ the tortoise coordinate, and leave the Êstudy of horizon-penetrating coordinates to future work. 

The angular coordinate also needs to be transformed to cure the infinite twisting near the horizon \cite{Teukolsky:1973ha, Krivan:1997hc}. Further, a rescaling  respecting the fall-off behavior of the curvature perturbations is necessary. Finally, we separate the azimuthal dependence to arrive at a $2+1$ dimensional system. These transformations are listed as follows:
\begin{enumerate}
\item Introduce the tortoise coordinate $r_*$,
\be \label{tortoise} \frac{dr_*}{dr} = \frac{r^2+a^2}{\Delta}.\ee
\item Introduce the azimuthal coordinate $\tilde{\varphi}$,
\[ d\tilde{\varphi} = d\varphi+\frac{a}{\triangle}dr.\]
\item Rescale the curvature perturbation of spin weight $s$ according to its asymptotic fall-off behavior,
\[ \Psi = r^{-(2s+1)} \psi\,.\]
\item Separate the azimuthal dependence using the mode number $k$,
\[\psi_k = e^{i k \tilde{\varphi}} \psi.\]
\end{enumerate}

In the following, we drop the subscript $k$ from $\psi_k$ for conciseness of notation. The Teukolsky equation becomes \cite{PazosAvalos:2004rp}
\begin{widetext}
\begin{eqnarray} \label{eq:tk}
D^2\partial_t^2\psi &=& 
\frac{(r^2+a^2)^2}{\triangle} \partial_{r_*}^2\psi+\frac{1}{\sin\theta}\partial_\theta(\sin\theta\partial_\theta \psi) + \nonumber \\
&& - \frac{2}{\triangle}\left(Ms(a^2-r^2) + rs\triangle + i(as\triangle\cos\theta+2Mark)\right)\partial_t\psi + \nonumber \\
&&+ \frac{1}{r \triangle}\left( \triangle(8r^2+6a^2) + 2rs(r^2+a^2)(r-M) + 2iakr(r^2+a^2) \right)\partial_{r_*} \psi + \nonumber \\
&&+ \frac{1}{r^2 \triangle} \left(6\triangle^2 - r\triangle\left(6M(s+1)-r (7s+6) + \frac{r}{\sin^2\theta}(k+s\cos\theta)^2\right) + 2i akr(2rs (r-M)+3\triangle)\right)\,\psi.
\end{eqnarray}
\end{widetext}

These are standard transformations for numerical computations. Hyperboloidal compactification adds only two additional transformations to this list: a transformation of the time coordinate, and a compactification of the radial coordinate. 

\section{A hyperboloidal layer for the Teukolsky equation}

We present the hyperboloidal layer method in the language of standard transformations to make its application as straightforward as possible for future studies.

\subsection{Hyperboloidal compactification}
A hyperboloidal surface is an everywhere spacelike surface that approaches null infinity \cite{Friedrich83}. It has been known already in the early days of numerical relativity that such surfaces are beneficial for studying gravitational radiation \cite{Eardley:1978tr, Smarr:1977uf, Brill80, Gowdy81}. A systematic study of the hyperboloidal initial value problem started with Friedrich's work showing that future null infinity is smooth for certain nontrivial classes of initial data \cite{Friedrich86}. There has been substantial effort to numerically simulate such data using Friedrich's conformally regular field equations \cite{Frauendiener:2000mk, Husa:2002zc}.

Even though the benefits of hyperboloidal surfaces for radiation problems beyond the Einstein equations have been clear, it took more than two decades before the hyperboloidal method could be applied successfully to systems other than the conformally regular field equations. The first application of this kind is the study of magnetic monopoles by Fodor and R\'acz \cite{Fodor:2003yg, Fodor:2006ue}. Their hyperboloidal foliation is based on a scri-fixing gauge in Minkowski spacetime first explicitly constructed by Moncrief \cite{Moncrief00} (see also \cite{Husa:2002zc}). It was suggested that scri-fixing gauges should be beneficial also for  studies on black hole spacetimes, however, the construction ofÊ a good hyperboloidal coordinates turned out to be difficult \cite{Gentle:2000aq, Gowdy:2001ij, Andersson02a, Schmidt02, Calabrese:2005rs, vanMeter:2006mv}. 

The general construction of suitable hyperboloidal, scri-fixing gauges has been presented in \cite{Zenginoglu:2007jw}. We follow \cite{Zenginoglu:2007jw} to present the hyperboloidal compactification. The technique consists of introducing a compactifying radial coordinate $\rho$ and aÊ suitable time coordinate $\tau$.

\subsubsection{Radial compactification}
We map the infinite positive domain in the tortoise coordinate $r_*$ to a finite domain in the compactifying coordinate $\rho$ to compute the solution in an unbounded physical domain using a finite numerical grid. After compactification, the outer boundary of the numerical grid corresponds to infinity with respect to the physical coordinate. Infinity has a rich structure in spacetime. What part of infinity is included in the computational domain with radial compactification depends on the nature of the time surfaces, as discussed in the next subsection. 

The radial compactification can be performed conveniently with the following definition of a compactifying coordinate:
 \be\label{rho} r_* = \frac{\rho}{\Omega(\rho)}. \ee
The choice of $\Omega$ determines the properties of the compactification. Its zero set corresponds to infinity with respect to $r_*$. For example, a common choice encountered in the literature is $\Omega=1-\rho$. We need more freedom in the choice of compactification, therefore we will state the general properties that the function $\Omega$ needs to satisfy. These properties are those that are satisfied by a conformal factor in the Penrose conformal compactification picture  \cite{Penrose63,Penrose65}. 
For all finite $r_*$ we require $\Omega>0$. Denoting the zero set of $\Omega$ as $S$ (the $\rho$-coordinate location of $r_*$-infinity), we require
\[  \Omega(S) = 0, \quad \Omega'(S) \ne 0.  \]
where $\Omega'\equiv d\Omega/d\rho$. 

It is known that compactifying the radial coordinate along Cauchy surfaces results in resolution problems \cite{Orszag}. These problems can be avoided by a hyperboloidal time transformation that changes the compactification at spatial infinity to a compactification at null infinity.

\subsubsection{Time transformation}
A suitable time transformation shall leave the exterior timelike Killing field invariant so that no gauge dynamics is introduced into the stationary background. The new time coordinate $\tau$ satisfies then $\partial_\tau = \partial_t$, which is achieved by a transformation of the form $\tau = t - h(x^i)$, where $x^i$ are the spatial coordinates. The function $h$ is called the height function. Most useful foliations of stationary spacetimes are given via such a relation. We are interested in radially outgoing gravitational waves, therefore we choose the height function to depend only on the tortoise coordinate $r_*$
\be\label{tau} \tau = t - h(r_*). \ee
The height function is chosen such that the surfaces $\tau=\rm{const.}$ are hyperboloidal \cite{Zenginoglu:2007jw}. The specific choice we use is presented in the next section (see Eq.~\eqref{height}).

In summary, hyperboloidal compactification consists of two simple transformations that can be written in the form \eqref{rho} and \eqref{tau} with free functions $\Omega$ and $h$. The specific choice of these free functions depends on the background spacetime and its coordinatization. There is a large freedom in their choice which allows us to use standard coordinates in an interior domain as described below.

\subsection{Hyperboloidal layer for Kerr spacetime}
An essential advantage of the hyperboloidal method is that it requires modifications only in the asymptotic domain, which allows us to use arbitrary coordinates in the strong field regime. 

A transition zone is used in  \cite{Zenginoglu:2007jw} for matching arbitrary coordinates near the central black hole to hyperboloidal scri-fixing coordinates in the asymptotic domain. Subsequent numerical experiments showed that the transition zone requires the adjustment of many arbitrary parameters and leads to a sudden drop of outgoing characteristic speeds decreasing the efficiency of the hyperboloidal method \cite{Zenginoglu:2009ey, Zenginoglu:2009hd, Vega:2011wf}. 

A new technique that avoids these deficiencies is the hyperboloidal layer presented in \cite{Zenginoglu:2010cq}. The idea is to attach a radial shell to the truncated computational domain in which the foliation is hyperboloidal and the radial coordinate is compactified. The layer needs to be attached in a sufficiently smooth way for the stability of the numerical computation.

We would like to have the new coordinate $\rho$ agree with the tortoise coordinate in the strong field region. Denoting the location of the interface between the interior domain and the layer by $R_*$, we set $\Omega=1$ for $\rho<R_*$ so that $\rho\equiv r_*$ in the interior. We also require that the coordinates agree up to a certain order at the interface. These conditions are fulfilled by the following expression \cite{Bernuzzi:2011aj, Zenginoglu:2010cq}
\be\label{omega} \Omega = 1- \left(\frac{\rho-R_*}{S-R_*}\right)^4
\Theta(\rho-R_*)\,,\ee
where $\Theta$ denotes the step function and $S$ denotes the coordinate location of the outer boundary. We emphasize that various similar choices for the radial compactification are possible. 

The original construction of a height function for a hyperboloidal layer \cite{Zenginoglu:2010cq} requires unit characteristic speed in the layer with respect to the layer coordinates $\tau$ and $\rho$. An equivalent but more intuitive requirement is that outgoing null rays have the same representation in the layer coordinates as in the interior coordinates \cite{Bernuzzi:2011aj}. In Schwarzschild spacetime with the tortoise coordinate the requirement reads $\tau-\rho = t-r_*$.  From this relation together with \eqref{tau} we get
\be\label{height} h = r_* - \rho(r_*) = \frac{\rho}{\Omega(\rho)} - \rho.\ee
In the domain where $\Omega=1$, we have $h=0$ so that we obtain the standard coordinates. For the $r_*$ derivative of the height function we get
\be\label{boost} H := \frac{dh}{dr_*}(\rho) = 1-\frac{d\rho}{dr_*} = 1 - \frac{\Omega^2}{\Omega - \rho\, \Omega'}.\ee
We refer to $H$ as the boost function. It vanishes in the interior domain where standard coordinates are used with $\Omega=1$; it is unity at infinity, where $\Omega=0$. 

The Refs.~\cite{Zenginoglu:2010cq, Bernuzzi:2011aj} use the prescription \eqref{boost} in Minkowski spacetime with standard coordinates, and in Schwarzschild spacetime with the tortoise coordinate. In this paper we apply it for the Kerr spacetime in Boyer--Lindquist tortoise coordinates. This procedure leads to a regular hyperboloidal compactification in Kerr spacetime, even though the outgoing null rays do not have the simple form $t-r_*$, because the Kerr metric in the tortoise Boyer-Lindquist coordinates has asymptotically the same form as the Schwarzschild metric in tortoise Schwarzschild coordinates, or the Minkowski metric in standard coordinates. This feature is another evidence for the flexibility of hyperboloidal coordinates. Only the asymptotic behavior of the null cone is relevant for the regularity of hyperboloidal compactification, whereas in the characteristic method the exact local form of null cone plays an essential role, thereby restricting the coordinate transformations \cite{Pretorius:1998sf, FletcherLun, Venter:2005cs, Bai:2007rs}.

The choices \eqref{rho} and \eqref{tau} with the free functions explicitly given in \eqref{omega} and \eqref{height} completely fix the coordinate transformation  from $\{t,r_*\}$ to $\{\tau,\rho\}$ that gives us a hyperboloidal layer in Kerr spacetime for the coordinate domain $[R_*,S]$.

\subsection{Hyperboloidal compactification of the Teukolsky equation}

The transformations discussed in the previous section can be regarded as additional items in the list of transformations for the numerical solution of the Teukolsky equation given in Section \ref{trafos}. 
\begin{enumerate}[resume]
\item Introduce a compactifying coordinate $\rho$, 
\be \label{eq:rho} r_*= \frac{\rho}{\Omega(\rho)}.\ee 
\item Introduce a time coordinate $\tau$,
\be \label{eq:tau} \tau = t-h(r_*).\ee 
\end{enumerate}
The compactification function $\Omega(\rho)$ and the height function $h(r_*)$ are determined via \eqref{omega} and \eqref{height}. With the coordinate transformations at hand, we can now proceed to transform the Teukolsky equation. 

It is a priori not clear that a simple coordinate transformation will lead to a regular hyperboloidal compactification of the Teukolsky equations. There are various possibilities to study the regularity of such compactification. One is to derive conformal Teukolsky equations with respect to a conformally extended, regular Kerr metric. It has been shown in a previous application of hyperboloidal compactification using the Teukolsky formalism in Schwarzschild spacetime (called the Bardeen--Press equation after \cite{Bardeen:1973xb}) that such an analysis is not necessary \cite{Zenginoglu:2008uc}. However, the method of \cite{Zenginoglu:2008uc} still requires the calculation of the Teukolsky equations ab initio in a general orthonormal Newman--Penrose frame which is then adapted to a hyperboloidal scri-fixing gauge to ensure regularity of hyperboloidal compactification. The method we use in this paper is easier to apply.

A general operator for an equation in the independent variables $\{t,r_*,\theta\}$, such as the Teukolsky equation \eqref{eq:tk}, can be written as $\mathcal{O} [\psi] = 0$, where $\mathcal{O}$ is the differential operator
\begin{eqnarray*} 
\mathcal{O} &=&  A^{tt} \partial_t^2 + A^{tr_*}\partial_t \partial_{r_*} + A^{r_* r_*}\partial_{r_*}^2 + A^{\theta\theta} \partial_\theta^2 \nonumber \\
&& + B^t \partial_t  +  B^{r_*} \partial_{r_*}  +  B^\theta \partial_\theta +  C.  
\end{eqnarray*}
The coefficients depend on $r_*$, and $\theta$; they can be read off from \eqref{eq:tk}. With the transformations \eqref{eq:rho} and \eqref{eq:tau} the derivative operators with respect to the old coordinates can be written in terms of the new coordinates as
\[  \partial_t = \partial_\tau, \qquad \partial_{r_*} = -H\partial_\tau + (1-H)\,\partial_\rho. \]
We used \eqref{boost} in the expression for $\partial_{r_*}$. 
We write the transformed operator as
\begin{eqnarray} \label{op_tk}
\mathcal{O} &=&   A^{\tau\tau} \partial_\tau^2 + A^{\tau\rho} \partial_\tau \partial_{\rho} + A^{\rho\rho} \partial_{\rho}^2 + A^{\theta\theta} \partial_\theta^2 \nonumber \\
&& + B^\tau \partial_\tau  +  B^{\rho} \partial_{\rho}  +  B^\theta \partial_\theta +  C,  
\end{eqnarray}
where
\begin{eqnarray*}
A^{\tau\tau} &=& A^{tt} - HA^{tr_*} + A^{r_*r_*}H^2, \\
A^{\tau\rho} &=& (1-H) (A^{tr_*}-2HA^{r_*r_*}), \\
A^{\rho\rho} &=& (1-H)^2 A^{r_* r_*}, \\
B^{\tau} &=& B^t - HB^{r_*} - A^{r_*r_*}H' (1-H),\\
B^\rho &=& (1-H) (B^{r_*} - A^{r_*r_*}H'),
\end{eqnarray*}
and $H':=dH/d\rho$. All coefficients are functions of $\rho$, and $\theta$.

The transformed system is by construction equivalent to the original system \eqref{eq:tk} in the interior domain $\rho<R_*$ where we have $H=0$. We need to make sure that the hyperboloidal compactification is regular at infinity where $H=1$. Note that asymptotically $(1-H)\sim \Omega^2 \sim r_*^{-2} \sim r^{-2}$. We see by inspection that the coefficients of the Teukolsky equation \eqref{eq:tk} have the correct asymptotic behavior ensuring regularity of the hyperboloidal compactification. The transformed coefficiencts have explicitly finite values at future null infinity. We mention that this feature is not special to the Teukolsky equations. Similar regular hyperboloidal compactifications for other wave equations have been studied as mentioned in the Introduction. 

We also need to make sure that the transformed system is pure outflow at the outer boundary so that no boundary conditions are required. This condition, along with the regularity, can be checked by evaluating the transformed equations at infinity, that is, at $\{\rho=S\}$. We obtain for $s=-2$
\begin{eqnarray*}
\bar{D} \partial_\tau^2 \psi &=& - S(S-R_*)\partial_\tau \partial_\rho \psi + 2 \partial_\theta^2 \psi \nonumber \\
&& - 4(4M+ i a (k-2\cos\theta)) \partial_\tau \psi \nonumber \\
&& + 2 \cot\theta \partial_\theta \psi - 2 ( 3+k^2-4k\cos\theta +\cos(2\theta)) \psi,
\end{eqnarray*}
where $\bar{D} = S(S-R_*) - 2 a^2 \sin^2\theta$. 
The principal part of the operator is of the form \mbox{$C_1\, \partial_\tau(\partial_\tau + \partial_\rho) + C_2 \, \partial_\tau^2+2\partial^2_\theta$}, where $C_1$ and $C_2$ are coefficients. Therefore, the modes of the system propagate either along the boundary at $\{\rho=S\}$, or out of the domain along the level sets of characteristics $\tau-\rho$. As expected, there are no incoming modes.

\section{Numerical Results}
In this section, we present tests using the hyperboloidal compactification of the Teukolsky equation. Our main result is the efficiency and accuracy of the hyperboloidal layer method providing unambiguous waveforms at future null infinity from large- and extreme-mass-ratio inspirals in Kerr spacetime.

\subsection{The numerical method}
For the discretization of the continuous equation we use a two-step Lax--Wendroff algorithm as in \cite{Krivan:1997hc}. After performing the transformations presented in the previous sections, we rewrite the vacuum equation in the form
\begin{eqnarray} \label{eq:final} 
\partial_\tau^2 \psi &=& \tilde{A}^{\tau \rho}\partial_\tau\partial_{\rho}\psi + \tilde{A}^{\rho \rho}\partial_{\rho}^2 \psi + \tilde{A}^{\theta\theta} \partial_\theta^2 \psi \nonumber \\
&+& \tilde{B}^\tau\partial_\tau\psi + \tilde{B}^{\rho} \partial_{\rho} \psi + \tilde{B}^\theta \partial_\theta \psi + \tilde{C} \psi, \end{eqnarray}
where the coefficients with a tilde are obtained by dividing the coefficients of the operator in \eqref{op_tk} by $-A^{\tau\tau}$. We put the equation \eqref{eq:final} in first-order form (in $\rho$ and $\tau$) by defining a new field variable 
\begin{eqnarray}
\pi &\equiv& \partial_{\tau}{\psi} + b \, \partial_{\rho}\psi \; , \\
b & \equiv & - (\tilde{A}^{\tau \rho} + \sqrt{ (\tilde{A}^{\tau \rho})^{2} + 4 \tilde{A}^{\rho \rho}})/2 \; .
\end{eqnarray}
We chose to use this first-order form of the equation for our numerical evolutions, because it has been discovered through extensive experimentation \cite{Krivan:1997hc} that such a form is ideally suited for long stable evolutions. Then equation \eqref{eq:final} with source terms \mbox{\boldmath{$T$}} takes the form 
\begin{eqnarray}
\label{eq:evln}
\partial_{\tau} \mbox{\boldmath{$u$}} + \mbox{\boldmath{$M$}} \partial_{\rho}\mbox{\boldmath{$u$}} 
+ \mbox{\boldmath{$Lu$}} + \mbox{\boldmath{$Au$}} = \mbox{\boldmath{$T$}},
\end{eqnarray}
where 
\begin{equation}
\mbox{\boldmath{$u$}}\equiv\{\psi_R,\psi_I,\pi_R,\pi_I\}
\end{equation}
is the solution vector. The subscripts $R$ and $I$ refer to the real and imaginary parts respectively. The matrices {\boldmath{$M$}}, {\boldmath{$A$}} and {\boldmath{$L$}} are obtained from \eqref{eq:final} as in \cite{Sundararajan:2007jg}. Here it will suffice to simply indicate the final form taken by these matrices: 
\begin{equation}
\mbox{\boldmath{$M$}} \equiv \left(\begin{matrix}
                    b  &   0   &  0     &  0 \cr
                    0  &   b   &  0     &  0 \cr
                    m_{31}  &   m_{32}  & -b  &  0 \cr
                    -m_{32}  &   m_{31} &  0  & -b \cr
                \end{matrix}\right) \; ,
\label{m_matrix}
\end{equation}
\begin{equation}
\mbox{\boldmath{$A$}} \equiv \left(\begin{matrix}
                    0  &   0   &  -1  &  0 \cr
                    0  &   0   &  0  &  -1 \cr
                    a_{31} & a_{32} & a_{33} & a_{34} \cr
                    -a_{32} & a_{31} & -a_{34} & a_{33} \cr
                \end{matrix}\right) \; ,
\label{a_matrix}
\end{equation}
and
\begin{equation}
 \mbox{\boldmath{$L$}} \equiv \left(\begin{matrix}
                    0  &   0   &  0  &  0 \cr
                    0  &   0   &  0  &  0 \cr
                    l_{31}  &   0   &  0  &  0 \cr
                    0  &   l_{31}   &  0  &  0 \cr
                \end{matrix}\right)\;.
\label{l_matrix}
\end{equation}
The angular derivatives are encoded in {\boldmath{$L$}}.

The main advantage of casting the equation \eqref{eq:final} into the form \eqref{eq:evln} is that the system has advantageous properties in the variable $\rho$. The matrix {\boldmath{$M$}} has a complete set of linearly independent eigenvectors with real eigenvalues. This is not a rigorous statement on the hyperbolicity of the system because the matrix {\boldmath{$L$}} contains second-order angular derivatives. Nevertheless, experiments show that the system is numerically well-behaved. A time-explicit evolution scheme is developed for this system of linear partial differential equations using the two-step, second-order Lax-Wendroff finite-difference method. We write \eqref{eq:evln} as 
\begin{equation}
\partial_{\tau} \mbox{\boldmath{$u$}} + \mbox{\boldmath{$D$}}
\partial_{\rho} \mbox{\boldmath{$u$}}
=  \mbox{\boldmath{$S$}}\; , 
\label{new_teu2}
\end{equation}
where
\begin{equation}
 \mbox{\boldmath{$D$}} \equiv \left(\begin{matrix}
                    b &   0   &  0  &  0 \cr
                    0  &   b   &  0  &  0 \cr
                    0  &   0   &  -b  &  0 \cr
                    0  &   0   &  0  &  -b \cr
                \end{matrix}\right),
\label{d_matrix}
\end{equation}
\begin{equation}
\mbox{\boldmath{$S$}} = -(\mbox{\boldmath{$M$}} - \mbox{\boldmath{$D$}})
\partial_{\rho}\mbox{\boldmath{$u$}}
- \mbox{\boldmath{$L$}}\mbox{\boldmath{$u$}} 
- \mbox{\boldmath{$A$}}\mbox{\boldmath{$u$}} + \mbox{\boldmath{$T$}} .
\end{equation}
Each iteration consists of two steps. In the first step, the solution vector between grid points is obtained from
\begin{eqnarray}
\label{lw1}
\mbox{\boldmath{$u$}}^{n+1/2}_{i+1/2} &=& 
\frac{1}{2} \left( \mbox{\boldmath{$u$}}^{n}_{i+1}
                  +\mbox{\boldmath{$u$}}^{n}_{i}\right)
- \\
&  &\frac{\delta \tau}{2}\,\left[\frac{1}{\delta \rho} \mbox{\boldmath{$D$}}^{n}_{i+1/2}
  \left(\mbox{\boldmath{$u$}}^{n}_{i+1}
                  -\mbox{\boldmath{$u$}}^{n}_{i}\right)
- \mbox{\boldmath{$S$}}^{n}_{i+1/2} \right] \; .\nonumber
\end{eqnarray}
This is used to compute the solution vector at the next time step,
\begin{equation}
\mbox{\boldmath{$u$}}^{n+1}_{i} = 
\mbox{\boldmath{$u$}}^{n}_{i}
- \delta \tau\, \left[\frac{1}{\delta \rho} \mbox{\boldmath{$D$}}^{n+1/2}_{i}
  \left(\mbox{\boldmath{$u$}}^{n+1/2}_{i+1/2}
                  -\mbox{\boldmath{$u$}}^{n+1/2}_{i-1/2}\right)
- \mbox{\boldmath{$S$}}^{n+1/2}_{i} \right] \, .
\label{lw2}
\end{equation}
The angular subscripts are dropped in the above equation for conciseness. All angular derivatives are computed using second-order, centered finite difference expressions. Symmetries of the spheroidal harmonics are used to determine the angular boundary conditions: For even $|m|$ modes we have $\partial_\theta\psi =0$ at $\theta = 0,\pi$; for odd $|m|$ modes we have $\psi =0$ at $\theta = 0,\pi$. We impose ``ingoing'' boundary conditions at the inner radial boundary. We do not need to apply any boundary condition at the outer boundary.

The small compact object inspiraling into the central supermassive black hole is modeled as a point particle in the large- or extreme-mass-ratio limit. The representation of point particle in mathematics is performed typically by a Dirac delta distribution. More specifically, the particle energy-momentum tensor takes the form 
\begin{eqnarray}
T_{\alpha\beta} &=& \mu \frac{u_\alpha u_\beta}{\Sigma \dot \tau \sin\theta}
\delta\left[r_* - r_*(\tau)\right]
\delta\left[\theta - \theta(\tau)\right]
\delta\left[\phi - \phi(\tau)\right]\;.
\nonumber 
\end{eqnarray}
where $u_\alpha$ and $\mu$ refer to the 4-velocity and the rest mass of the particle, respectively. Note that $\dot \tau \equiv d \tau /ds$ appears in the denominator of this expression, which tends to $\infty$ as the particle approaches the horizon (we denote by $s$ the proper time along the particle's trajectory). Thus, the particle source-term on the right-hand-side of \eqref{eq:evln} smoothly decays away as the particle approaches the horizon, thereby allowing the evolution equation to gradually transition into its homogeneous form. This source-term is constructed using the energy-momentum tensor describing a point particle moving in Kerr space-time depicted above. The explicit expression
for {\boldmath{$T$}} is very lengthy and not particularly illuminating. Here, we simply point out that the final expression is built using Dirac delta functions in $r$ and $\theta$, as well as first and second derivatives of the delta functions in these variables. These terms have coefficients that are complex functions of the black hole's physical parameters and also the trajectory of the point particle. Details on the particle source-term, along with the representation of the delta function and its derivatives on a numerical grid, are given in \cite{Burko:2006ua, Sundararajan:2007jg, Sundararajan:2008zm, Sundararajan:2010sr}.  

The trajectory of the particle in a decaying orbit around the central black hole is computed separately and is then used to calculate the source-term mentioned above, which in turn is fed into the Teukolsky equation solver code. The computation of the decaying trajectory can be broken into three distinct pieces: An early time {\em adiabatic} inspiral, in which the particle is approximated as evolving through a sequence of bound orbits of the central black hole that are calculated using an ``energy balance'' approach; a late time geodesic {\em plunge}, in which the particle falls into the black hole and radiation reaction is ignored; and an intermediate {\em transition} regime that smoothly connects these two pieces. Details on how these steps are handled and used by the Teukolsky equation solver code can be found in \cite{Sundararajan:2010sr}. It is straightforward to make use of decaying orbital trajectories from other approaches, such as the EOB or the self-consistent self-force approaches. 

\subsection{Waveforms}

After solving the Teukolsky equation with a particle source in the time-domain, we compute the gravitational waveforms $h_+$ and $h_\times$ using the simple relationship shown below that is valid in the far field, 
\begin{equation}
\psi \to \frac{r}{2}\left(\frac{\partial^2 h_+}{\partial \tau^2} - i
\frac{\partial^2 h_\times}{\partial \tau^2}\right)\;.
\label{eq:psi4hphm}
\end{equation}
Note that in our study we have direct access to the far field because null infinity is part of the computational domain. This allows us to use the above relation as an equality and to extract the waveform cleanly. 

\begin{figure}[ht]
\center
\includegraphics[width=0.47\textwidth]{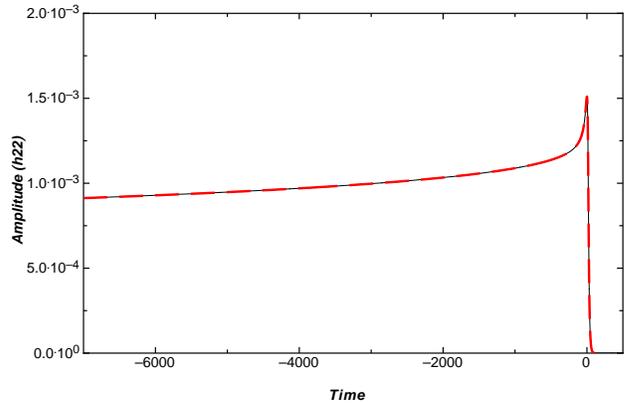}\hfill
\caption{The full waveform amplitude emitted by the inspiral, plunge, and ringdown of the particle into the central Kerr black hole calculated both with the hyperboloidal code (black), and the Cauchy code (dashed, red). The merger time ($\tau=0$) is defined rather arbitrarily by the time of maximum amplitude. \label{waveform}}
\end{figure}

\begin{figure}[ht]
\center
\includegraphics[width=0.47\textwidth]{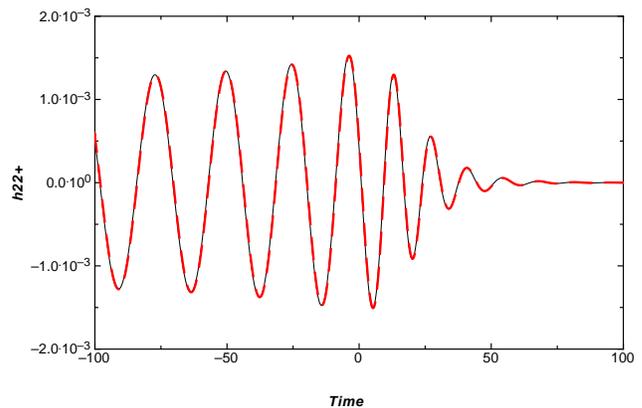}
\caption{The waveform emitted during the final plunge phase of the EMRI inspiral with the largest gravitational radiation emission.\label{waveform_zoom}}
\end{figure}

Figures \ref{waveform} and \ref{waveform_zoom} depict the gravitational waveform $h_{\ell m}$ measured at future null infinity, with $\ell = m = 2$. The mass-ratio for this evolved binary system is $\mu = 10^{-3}$ and the spin of the central Kerr black hole is $a = 0.5$. The system undergoes a decay due to gravitational wave emission over approximately $123$ full cycles. The computational domain is $\rho \in [-50, 50]$ and $\theta \in [0, \pi]$ with grid sizes of $3125$ and $32$ respectively. The hyperboloidal layer starts at $\rho = 14$, therefore the particle's orbit never crosses into the layer. There are a total number of $562,500$ time-steps in this computation. 

\begin{figure}[ht]
\center
\includegraphics[width=0.4\textwidth]{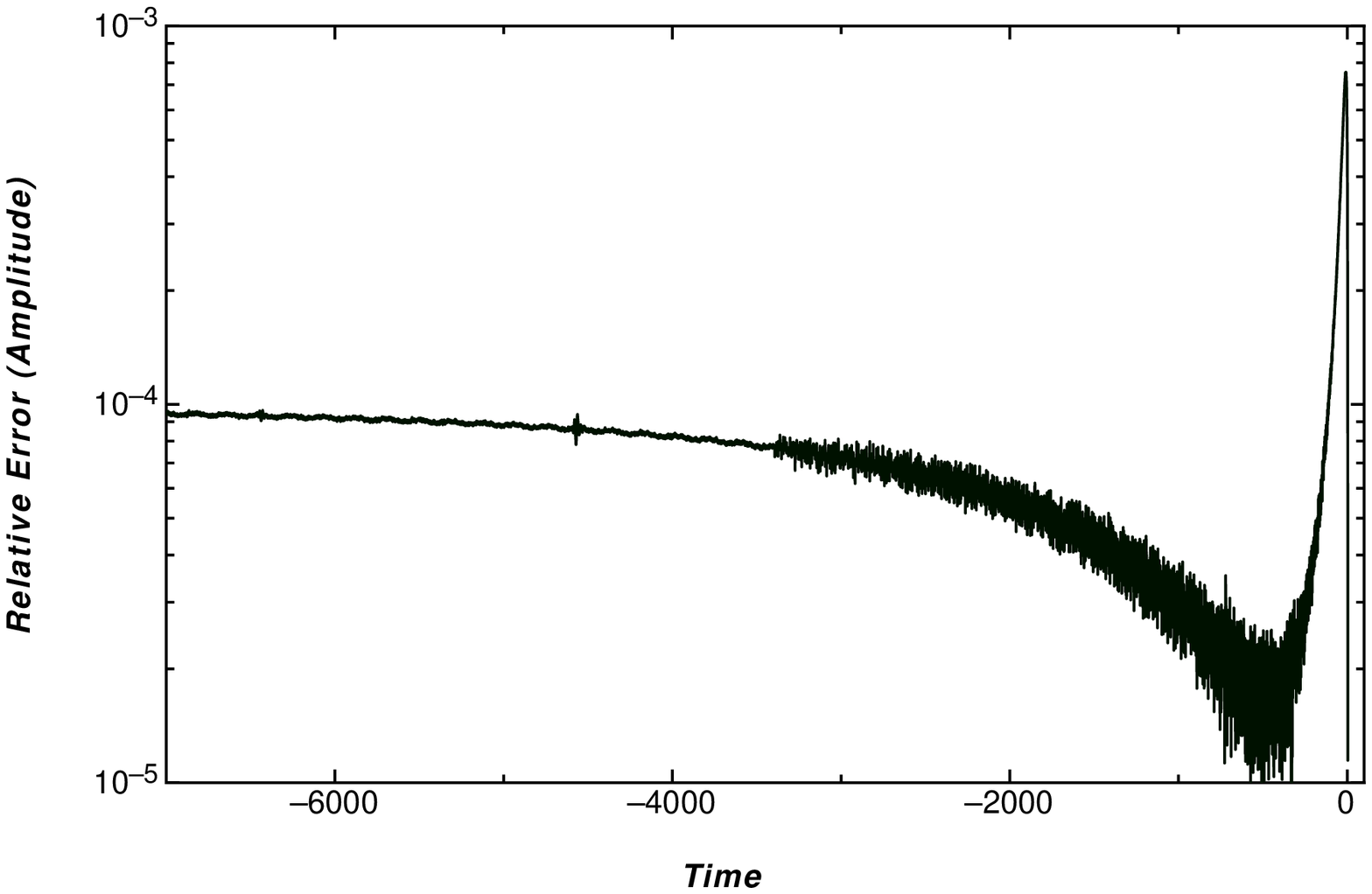}\hfill
\includegraphics[width=0.4\textwidth]{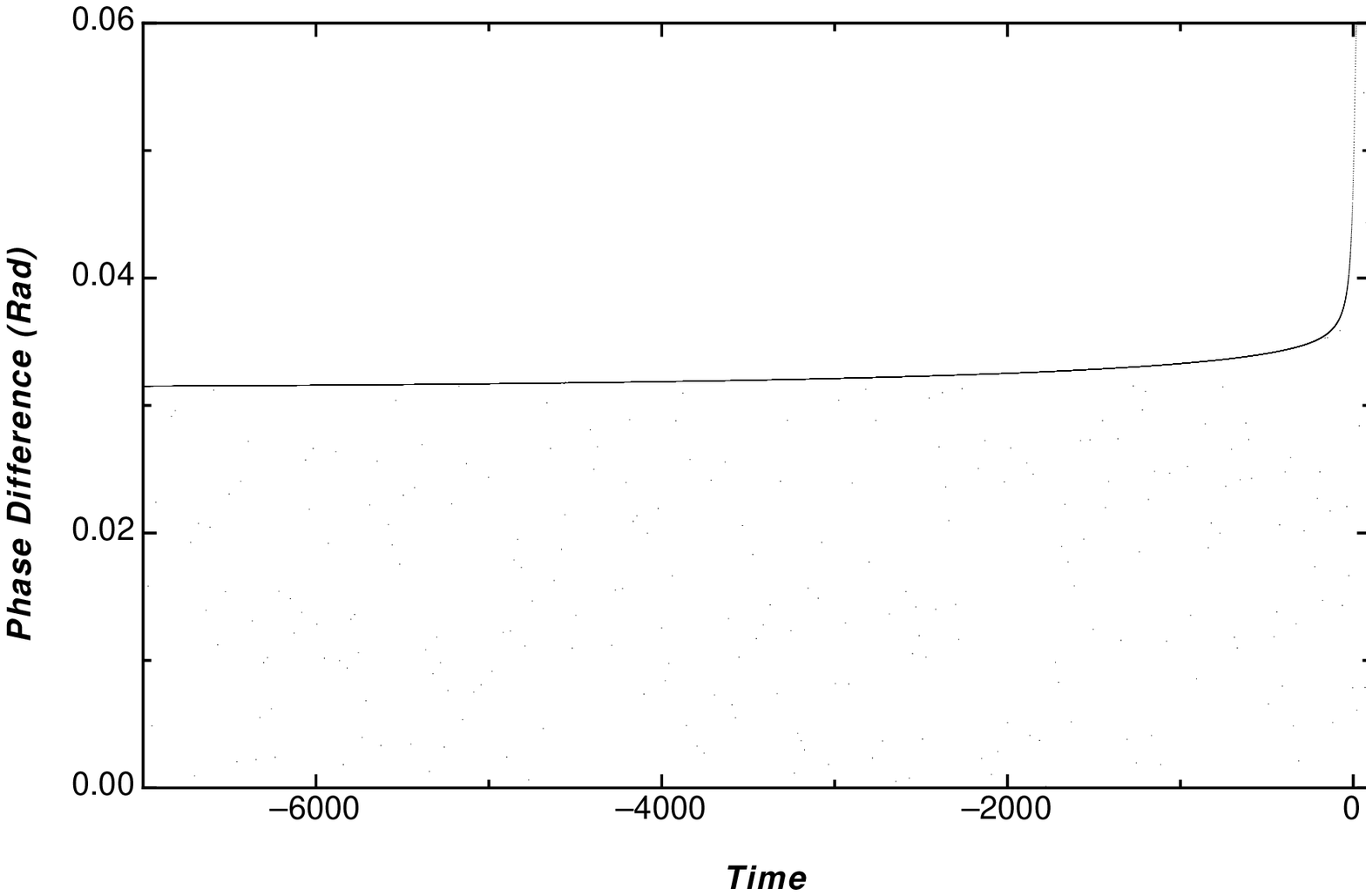}
\caption{Relative difference in amplitude (top) and phase difference (bottom) between the hyperboloidal code and the Cauchy code. For clarity, it is worth pointing out that the phase difference reaches a maximum in the ballpark of $0.06$ rad near $\tau=0$.\label{differences}}
\end{figure}

The figures depict the waveforms from both the original Cauchy code \cite{Burko:2006ua, Sundararajan:2007jg} and the new code with the hyperboloidal layer in Kerr spacetime developed in this work. The Cauchy code approaches spatial infinity, and therefore cannot provide direct access to waveforms at null infinity. To obtain data from the Cauchy code to compare with that from the hyperboloidal layer code, we extract the waveforms at multiple radii on the spatial grid and extrapolate them along fixed values of retarded time, to infinity, using a simple second order polynomial expansion. We do not employ higher order extrapolation for the Cauchy code because the truncation error becomes larger than extrapolation error with the given resolution and second order finite differencing. Nevertheless, the waveforms from the two codes agree so well that it is difficult to make out any difference between the two in these figures. In Figure \ref{differences} we show the differences in the waveform amplitude and phase for the same data. The relative difference between the amplitudes is at the level of $10^{-3}$ and the maximum difference in phase is $0.06$ rad. A better agreement between extrapolated and null infinity waveforms can be obtained with more accurate extrapolation algorithms and higher-order methods in spherical symmetry \cite{Bernuzzi:2011aj}. 

\subsection{Fluxes}

Gravitational waves carry energy away from a binary system thus causing it to inspiral. The calculation of the energy carried away by these waves uses the standard expression,
\begin{equation}
\frac{dE}{d\tau} = \lim_{r\rightarrow\infty}\frac{1}{4\pi}
\left|\int_{-\infty}^\tau \psi\,dt'\right|^2\,d\Omega\;,
\label{eq:Edot}
\end{equation}
in a postprocessing step after time evolution. In standard computations, the limit to infinite radius in the above equation is approximated by application of the formula at finite but large radii, and a subsequent extrapolation of the corresponding values to infinity. One major advantage of the hyperboloidal method is that the limit in Eq.~\eqref{eq:Edot} can be replaced by local evaluation of the integral along future null infinity, which is the boundary of the computational domain. Therefore, the extrapolation of fluxes is not necessary, and the energy can be computed cleanly. There are more subtleties in the computation of the fluxes related to the integration in time from the infinite past \cite{Reisswig:2010di}. We ignore these and set the integration constant to zero. We observe empirically that this procedure does not lead to large errors. 

In Table \ref{fluxes} we show numerical values of the energy fluxes computed at null infinity by the new hyperboloidal layer code for several circular-equatorial orbits. Comparing these values with those from very high-accuracy frequency-domain approaches \cite{Finn:2000sy}, we note an agreement at the $99.95\%$ level. We interpret this high level of agreement with the frequency-domain fluxes as strong evidence for the accuracy of the code, especially as compared with the original time-domain Cauchy code that achieves agreement with frequency-domain fluxes at the level of $99\%$ \cite{Burko:2006ua, Sundararajan:2007jg}.  

\begin{table}
  \centering

    \begin{tabular}{|c|c|c|c|c|c|} \hline
  {\bf a} & {\bf Radius} & {\bf TD } & {\bf FD } & {\bf \% Diff.} & {\bf Conv.}\cr
   \hline \hline
   -0.5 & 10.576 &   4.5560    &  4.5584  & 0.053 & 1.96 \cr \hline
   +0.0 & 6.0000 &   73.683    &  73.720  & 0.050 & 2.05 \cr \hline
   +0.5 & 4.2753 &   284.72    &  284.88  & 0.056 & 2.04 \cr \hline
   +0.8 & 11.627 &   2.2272    &  2.2287  & 0.067 & 2.04 \cr \hline 
 \end{tabular}

\caption{This table depicts the numerical values of energy fluxes  for the $|m| = 2$ mode from various circular-equatorial orbits, computed by the time-domain hyperboloidal code (TD) and a high-accuracy frequency-domain (FD) code. The flux values are scaled up with the square of the inverse mass-ratio ($\mu^{-2}$) and by a factor of $10^5$. Convergence rate is also presented. }

\label{fluxes}
\end{table}

In Table \ref{fluxes}, we also show the standard three level convergence rates. We obtain clear second-order convergence, as expected. The fluxes depicted in the table were obtained via Richardson extrapolation using the data from the different grid resolutions. The computational domain is $\rho \in [-50, 50]$ and $\theta \in [0, \pi]$ with grid sizes $1250 \times 32$, $2500 \times 64$ and $5000 \times 128$. The hyperboloidal layer starts at $\rho = 14$, therefore these orbits never cross into that layer throughout the simulations. 

\begin{figure}[ht]
\center
\includegraphics[width=0.47\textwidth]{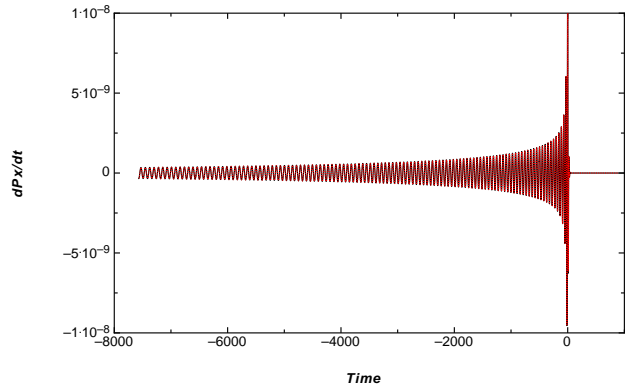}\hfill
\caption{Linear momentum flux measured at $\{\rho=S\}$ with the hyperboloidal code (black), and extrapolated to infinity with the Cauchy code (dashed, red). \label{mom_flux}}
\end{figure}

\begin{figure}[ht]
\center
\includegraphics[width=0.47\textwidth]{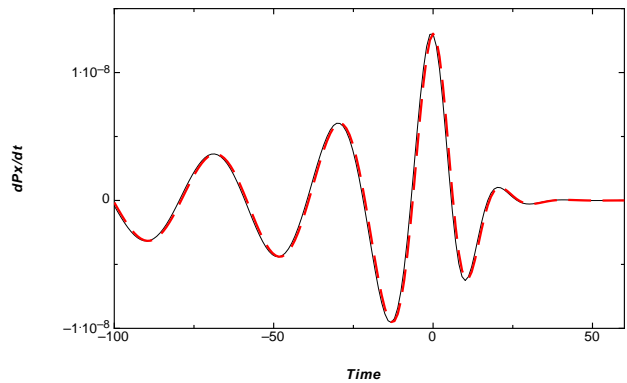}
\caption{Linear momentum flux as in Fig.~\ref{mom_flux} zoomed into the final plunge stage of the inspiral. \label{mom_flux_zoom}}
\end{figure}

\begin{figure}[ht]
\center
\includegraphics[width=0.47\textwidth]{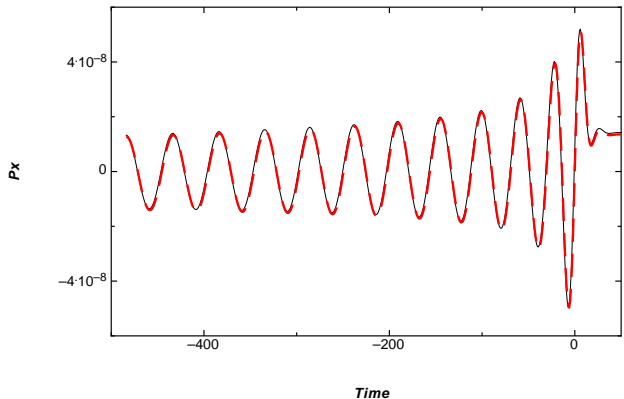}
\caption{Linear momentum in $x$ direction. \label{recoil}}
\end{figure}

\subsection{Kick velocities}
The asymmetric emission of gravitational waves during the inspiral of the small black hole into the supermassive central black hole leads to a recoil due to conservation of momentum. We compute this recoil by first computing the linear momentum flux carried away by the gravitation waves, and then performing a simple time-integral of this flux \cite{Sundararajan:2010sr}. To compute the momentum flux, we use the expression
\begin{equation}
\frac{dP^i}{d \tau} = \lim_{r\rightarrow\infty}\frac{1}{4\pi}\int n^i
\left|\int_{-\infty}^\tau \psi\,dt'\right|^2\,d\Omega\;.
\end{equation}
where the $n^i$ is the unit direction vector in spherical coordinates. As for Eq.~\eqref{eq:Edot}, the limit to infinity is replaced by local evaluation of the integral at future null infinity, which allows us to read off the momentum flux without an approximation in the radial direction. The integration constant resulting from starting the integration at $\tau=0$ is set arbitrarily to zero.

In Figures \ref{mom_flux}, \ref{mom_flux_zoom} and \ref{recoil} we depict the results from the computation of linear momentum flux and its time-integral. The mass-ratio for this computation is $\mu=10^{-3}$ with the central black hole spin $a = 0.7$. The computational domain is $\rho \in [-50, 50]$ and $\theta \in [0, \pi]$ with grid size $3125\times32$. The hyperboloidal layer starts at $\rho = 14$, therefore the particle's orbit never crosses into that layer throughout the simulation. There are  $2,250,000$ time-steps in this computation. Once again we superimpose results from the Cauchy code and the new hyperboloidal code which agree to a very high level. The final values of the recoil speed from the two different codes,  $1.645 \times 10^{-8}$ and $1.641 \times 10^{-8}$, agree to three significant digits, consistent with our previous results.  

\subsection{Efficiency} \label{sec:eff}

Previous sections provided evidence for the accuracy of the hyperboloidal method. The gain in accuracy in observables, such as the waveforms or energy fluxes, is mainly due to the direct numerical access to asymptotic quantities at future null infinity, which is part of the computational domain. The difference in those quantities, however, is rather small, essentially because extrapolation in given background spacetimes already gives reliable results. The basic advantage of the hyperboloidal method in this context is that it simplifies the accurate extraction of physical observables from numerical simulations: It eliminates the extrapolations.

The remarkable feature of the hyperboloidal method is that it provides improved accuracy at \emph{negative cost}. The numerical computation is much more efficient along hyperboloidal surfaces than along Cauchy surfaces. One may think that the calculation of waveforms at null infinity introduces additional computational expense for numerical simulations. In fact, the opposite is true: One gains \emph{more} accuracy for \emph{less} computational cost. 

The explanation of this property of hyperboloidal evolutions is that the standard method of using truncated Cauchy foliations with artificial outer boundary conditions is unnatural to study radiative solutions. An efficient computation of the asymptotic gravitational wave signal can only be achieved if the surfaces, along which the solution is computed, follow the outgoing signal closely. Gravitational waves propagate to infinity along null rays. Null rays and Cauchy surfaces diverge infinitely from each other in the asymptotic region. Cauchy surfaces are therefore inadequate for computing outgoing asymptotic radiation. Further, their truncation at a finite distance introduces errors through artificial outer boundary data contaminating the computation. 

To avoid such contamination in a standard Cauchy code the outer boundary is usually chosen to be causally disconnected from the region of interest. This implies that the duration of the simulation is limited by the size of the numerical grid. When null infinity is part of the computational domain, however, no such restriction applies. Therefore, in principle, we can perform arbitrary long evolutions, only restricted by the accumulation of the numerical truncation and the physical approximation errors. The efficiency of a hyperboloidal code as compared to a Cauchy-type code increases with the duration of the simulation. 

\begin{figure}[ht]
\center
\includegraphics[width=0.47\textwidth]{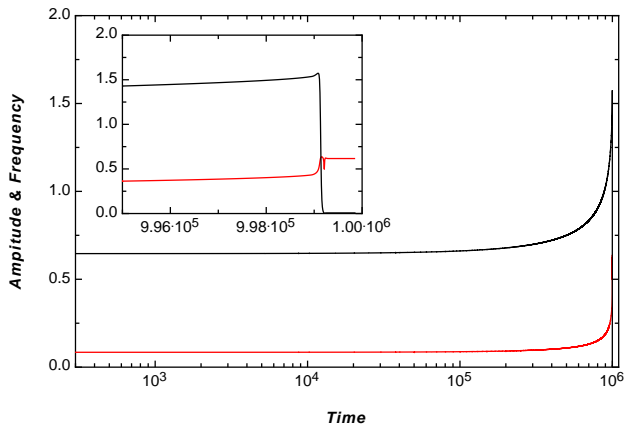}
\caption{An evolution that runs a million M. The frequency is depicted in red. The amplitude (top black line) has been scaled up by $1/\mu$. The inset shows the final merger.
\label{fig:verylong}}
\end{figure}

To demonstrate the efficiency of our hyperboloidal Teukolsky code, we performed an evolution of an EMRI that lasts for one million $M$. The spin of the central black hole is $0.9$, the mass ratio is $10^{-4}$. There are about $10,111$ orbital cycles in this simulation. The computational domain is $\rho \in [-50, 50]$ and $\theta \in [0, \pi]$ with the grid size $5000\times32$. The hyperboloidal layer starts at $\rho = 14$, therefore the particle's orbit never crosses into that layer throughout the simulation. There are over a $100$ million time-steps involved in this computation. In Fig.~\ref{fig:verylong} we show the amplitude and frequency against time in a half-logarithmic scale. The amplitude and frequency of the gravitational waveform at null infinity is fairly constant over most of the simulation, as expected. The main content of this plot is to demonstrate that such simulations in time-domain are now possible with the new method. Using the Cauchy code for such a simulation, would have required us to place the outer boundary at least as far as $500,000M$. So the Cauchy code would take $5000$ times longer to perform this simulation with equivalent local grid resolution. 
This number can be increased even further using horizon-penetrating coordinates.

A million $M$ in geometric units for a supermassive black hole of, say, $6$ million solar masses corresponds to about one Earth year in real time. The simulation for Fig.~\ref{fig:verylong} has been performed on a computer cluster using on the order of 1000 processor-cores in just under a day. This shows that we can perform long-time simulations of EMRIs much faster than the expected real time  duration of the events, and that large parameter studies in perturbation theory are computationally feasible with time-domain codes.

\section{Discussion and Outlook}

We presented the first numerical evolution scheme for computing gravitational perturbations of a rotating black hole spacetime including future null infinity.  We tested our method on the inspiral of a point particle into a central rotating black hole. The main advantages of the method are: (i) increase in accuracy by direct radiation extraction at future null infinity; (ii) a clean solution to the outer boundary problem in Kerr spacetime; and (iii) a great gain in efficiency for long time simulations.

Our method is based on the hyperboloidal compactification of Kerr spacetime presented in \cite{Zenginoglu:2007jw}. We also showed, for the first time, how the hyperboloidal layer developed in \cite{Zenginoglu:2010cq} can be employed in a Kerr spacetime. The layer technique allowes us to use arbitrary coordinates in an interior compact domain including the central black hole and the orbit of the particle, while compactifying the asymptotic domain along a hyperboloidal foliation. Because the hyperboloidal layer is attached to the interior domain sufficiently smoothly, only minor modifications of existing computational  infrastructure is needed (see \cite{Bernuzzi:2011aj} for a related study in Schwarzschild spacetime)

We showed that the hyperboloidal compactification of the Teukolsky equations can be performed by a simple coordinate transformation. There is no need for studying a conformally regular Kerr spacetime, or for calculating the equations ab initio in a general orthonormal Newman--Penrose frame as has been done in \cite{Zenginoglu:2008uc}. The coordinate transformation of the equations as given in Boyer--Lindquist coordinates leads naturally to a regular hyperboloidal compactification. Only the asymptotic behavior of the null cone is relevant for the hyperboloidal method, which implies that the simple prescription devised in Minkowski and Schwarzschild spacetimes can be used in Kerr spacetime with respect to Boyer--Lindquist tortoise coordinates. This provides evidence for the flexibility of the hyperboloidal approach, as opposed to the characteristic approach which restricts the coordinates locally for compactification at future null infinity \cite{Pretorius:1998sf, FletcherLun, Venter:2005cs, Bai:2007rs}.

We compared our results to a previous study of EMRIs in Kerr spacetime \cite{Sundararajan:2010sr}. The emphasis in this paper is not on new physics but on the numerical accuracy and efficiency of the hyperboloidal method as applied in Kerr spacetime. The waveforms at null infinity differ only slightly from their previous computation obtained by extrapolation. We showed that the hyperboloidal method gives better accuracy in the energy fluxes at infinity. Our strongest case for efficiency is the simulation of an inspiral that lasts one million $M$ in geometric units, which would have taken about 5000 times longer using the standard method.

In the future this method should be applied further to physically interesting problems. One could, for example, look for the new ringdown frequency mode for perturbations of a Kerr black hole discovered by Mino and Brink \cite{Mino:2008at} (see also \cite{Zimmerman:2011dx}) or for transient resonances expected for certain orbits in Kerr spacetime \cite{Flanagan:2010cd}. Such studies require high accuracy, so the description of the particle's motion needs to be improved, for example by including conservative effects in the kludge approach, by reading off the source terms from an EOB description, or by computing the self-consistent evolution driven by the high-order in mass ratio self-force of the particle. 

It is clear that there is much room for development in the accurate computations of EMRIs. It would be interesting to compare the different approaches to the EMRI problem in the same setting, similar to comparative studies performed in nonlinear numerical relativity \cite{Aylott:2009ya}. Comparison of physical invariants on given background spacetimes is simpler, therefore such a study should be easier to perform in the perturbative setting than in the nonlinear one. 

On a numerical level, aside from the technical improvements such as the use of high-order finite differencing or multidomain pseudospectral methods, implementation of horizon-penetrating coordinates near the black hole should lead to a further improvement in efficiency. Horizon-penetrating, hyperbolodial surfaces have certain advantages in perturbation theory over the standard Boyer--Lindquist or Schwarzschild time surfaces that intersect at the bifurcation sphere and approach spatial infinity \cite{Zenginoglu:2011jz}. In our context, horizon-penetrating coordinates should allow us to calculate accurately the absorbed fluxes at the horizon via the ingoing radiation represented by the curvature component $\psi_0$. 

Our calculation of an inspiral that corresponds to roughly a year in real time (depending on the mass of the central supermassive black hole) but takes only about one day of computation is a strong evidence for the efficiency of our numerical infrastructure (see Section \ref{sec:eff}). Note that the situation is reversed when fully nonlinear Einstein equations are solved for equal-mass, stellar, binary black-hole mergers. There, a typical computation takes about a month, but a stellar black-hole merger takes only milliseconds in real time. Therefore, it is imperative that the efficiency of nonlinear Einstein codes is increased. Given the physical simplicity of an equal-mass binary black-hole configuration, a combination of technical improvements including the hyperboloidal method, along with analytical insight into the problem, should allow us to simulate such systems much more efficiently Êin  the near future. 

The application of the hyperboloidal method to generic computations with the fully nonlinear Einstein equations is an outstanding problem. Recently there has been studies on various aspects of the hyperboloidal approach such as initial data \cite{Buchman:2009ew}, or gauge conditions \cite{Ohme:2009gn}. There are also suggestions on the evolution problem that do not require explicit regularity of the equations at future null infinity \cite{Zenginoglu:2008pw,Moncrief:2008ie, Bardeen:2011ip}. The only successful numerical implementation of such a formalism is by 
Rinne in axisymmetry~\cite{Rinne:2009qx}, but no generic computation could be presented so far. We hope that the insight provided by studies in perturbation theory will help the application of the hyperboloidal method to nonlinear Einstein equations.

\acknowledgements

AZ thanks Sebastiano Bernuzzi, Michael Jasiulek, and Alessandro Nagar for discussions. AZ is supported by the NSF Grant PHY-1068881, and by a Sherman Fairchild Foundation grant to Caltech in Pasadena. GK acknowledges research support from NSF Grant Nos. PHY-0902026, CNS-0959382, PHY-1016906 and PHY-1135664, and AFOSR DURIP Grant No. FA9550-10-1-0354. Some of the data presented in this work was generated on the Air Force Research Laboratory CONDOR supercomputer. GK also acknowledges support from AFRL under CRADA No. 10-RI-CRADA-09.

\vspace{7mm}

\bibliography{refs20111212}
\bibliographystyle{unsrt}

\end{document}